\documentclass[iop]{emulateapj}
\newcommand{\be}{\begin{equation}}
\newcommand{\ee}{\end{equation}}
\newcommand{\doh}{$\rm{D/H}$}
\newcommand{\od}{O'D}
\newcommand{\accO}{'O} 

\usepackage{natbib}
\usepackage{enumerate}
\usepackage{color}
\usepackage{graphicx}

\begin{document}
\title{Exploring the Origins of Deuterium Enrichments in Solar Nebular Organics}
\shorttitle{}
    \shortauthors{Cleeves et al.}

 \author{L. Ilsedore Cleeves\altaffilmark{1,5}, Edwin A. Bergin\altaffilmark{2}, Conel M. O'D. Alexander\altaffilmark{3}, Fujun Du\altaffilmark{2}, Dawn Graninger\altaffilmark{1}, Karin I. {\"O}berg\altaffilmark{1} and Tim J. Harries\altaffilmark{4}}

\altaffiltext{1}{Harvard-Smithsonian Center for Astrophysics, 60 Garden Street, Cambridge, MA 02138}
\altaffiltext{2}{Department of Astronomy, University of Michigan, 1085 S. University Ave., Ann Arbor, MI 48109}
\altaffiltext{3}{DTM, Carnegie Institution of Washington, Washington, DC 20015}
\altaffiltext{4}{Department of Physics and Astronomy, University of Exeter, Stocker Road, Exeter, EX4 4QL, UK} 
\altaffiltext{5}{Hubble Fellow, ilse.cleeves@cfa.harvard.edu}

\begin{abstract}
Deuterium-to-hydrogen (D/H) enrichments in molecular species provide clues about their original formation environment.  The organic materials in primitive solar system bodies have generally higher D/H ratios and show greater D/H variation when compared to D/H in solar system water. We propose this difference arises at least in part due to 1) the availability of additional chemical fractionation pathways for organics beyond that for water, and 2) the higher volatility of key carbon reservoirs compared to oxygen. We test this hypothesis using detailed disk models, including a sophisticated, new disk ionization treatment with a low cosmic ray ionization rate, and find that disk chemistry leads to higher deuterium enrichment in organics compared to water, helped especially by fractionation via the precursors CH$_2$D$^+$/CH$_3^+$. We also find that the D/H ratio in individual species varies significantly depending on their particular formation pathways. For example, from $\sim20-40$~AU, CH$_4$ can reach $\rm{D/H\sim2\times10^{-3}}$, while D/H in CH$_3$OH remains locally unaltered. Finally, while the global organic D/H in our models can reproduce intermediately elevated D/H in the bulk hydrocarbon reservoir, our models are unable to reproduce the most deuterium-enriched organic materials in the solar system, and thus our model requires some inheritance from the cold interstellar medium from which the Sun formed. 
\end{abstract}


\section{Introduction} 
\begin{figure*}
\begin{centering}
\includegraphics[width=0.77\textwidth]{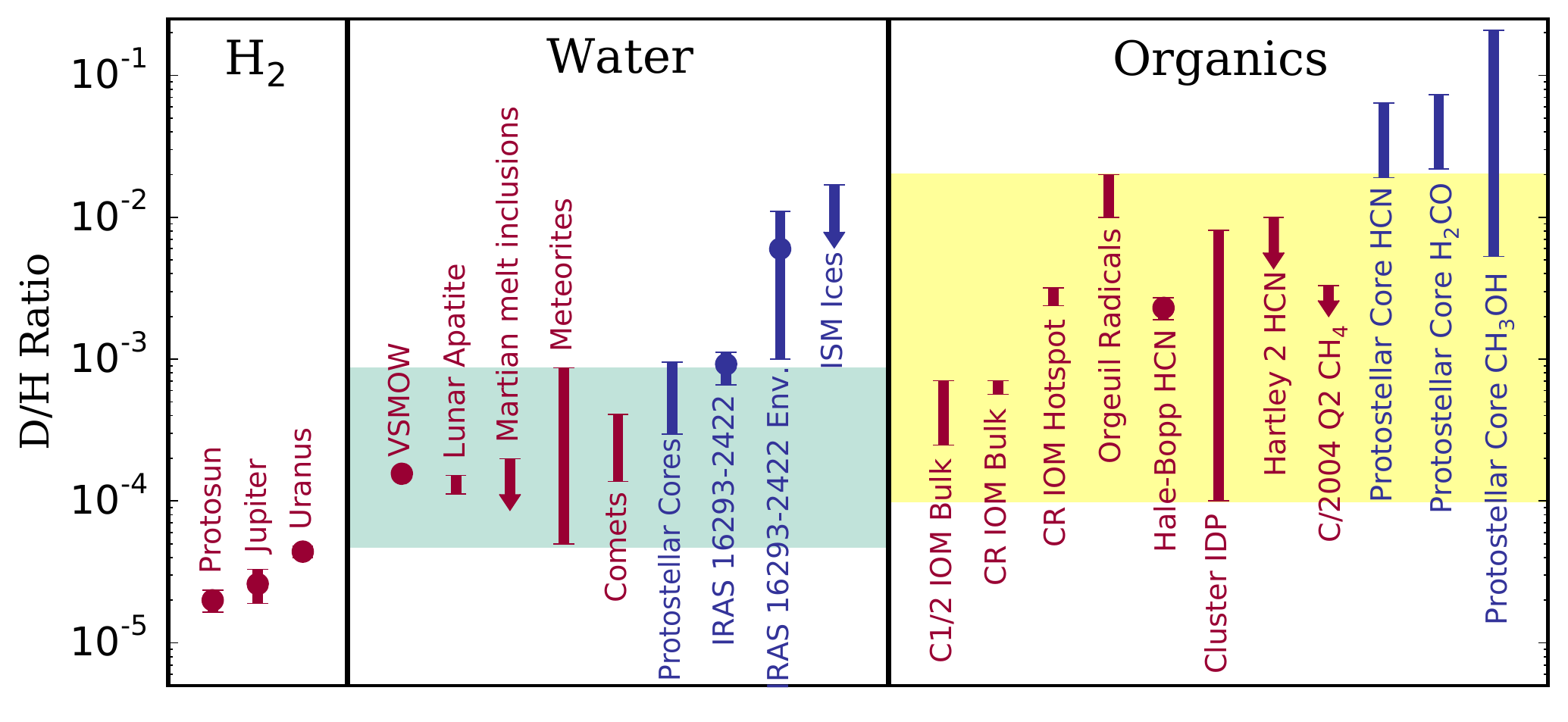}  
\caption{$\rm{\left[D/H\right]}$ in H$_2$, water and organics. Teal (yellow) band highlights the range of $\rm{\left[D/H\right]}$ in solar system water (organics). References from left to right: {\bf HD:} \citet{geiss2003}; \citet{mahaffy1998}; \citet{feuchtgruber2013}; {\bf H$_2$O:} \citet{coplen1994}; \citet{barnes2014}; \citet{usui2012}; \citet{alexander2012}; {\em Comets:} \citet{hartogh2011,eberhardt1995,bockeleemorvan1998,meier1998a}; {\em Interstellar:} \citet{persson2014},  \citet{persson2012}, \citet{parise2003}, \citet{coutens2012}; {\bf Organics:} \citet{alexander2007}; \citet{alexander2007}; \citet{busemann2006}; \citet{gourier2008}; \citet{eberhardt1995}; \citet{messenger2000};  \citet{milam2011}; \citet{kawakita2009}; \citet{roberts2002}; \citet{roberts2002}; \citet{parise2006}. \label{fig:dhplot}}
\end{centering}
\end{figure*}

The origin of Solar System water and organic materials is a key question of astrobiology. In cold environments exposed to ionizing radiation, the amount of deuterium relative to hydrogen (\doh) in both water and organics can become chemically enhanced. Thus, the \doh\ ratio encodes important information about the physical conditions where these molecules originally formed \citep[e.g.,][]{millar1989}.  Indeed, both water and organic material in the solar system and in the interstellar medium exhibit extreme enrichments in their \doh\ as compared to the bulk gas reservoir \citep[$\rm{\left[D/H\right]_{H_2}}=2.0\pm0.35\times10^{-5}$ during the epoch of planet formation;][]{geiss2003}, where interstellar enrichments are typically higher than those in primitive solar system bodies (see Figure~\ref{fig:dhplot}). There are two possible environments where these deuterium enrichments were originally implanted: (1) the cold interstellar medium, activated by galactic cosmic rays (CRs), and/or (2) the outermost regions of the protoplanetary disk in the presence of stellar or non-stellar ionization.  Therefore, (1) corresponds to an inheritance scenario, while (2) is a local (disk) synthesis process. Either of these scenarios has very different implications for the availability of these important life-fostering ingredients to young planets during their formation. 

Recently, \citet{cleeves2014wat}, hereafter Paper~I, found that pure disk synthesis for water was unable to reproduce the range of $\rm{\left[D/H\right]_{H_2O}}$ ratios among solar system bodies, including Earth's oceans (VSMOW), even though temperatures were sufficiently cold in large portions of the outer solar nebula.  Paper~I included a more realistic low CR ionization environment that was originally proposed in \citet{glassgold1997} \citep[see also][]{turner2009} and explored in detail by \citet{cleeves2013a}, and has been subsequently confirmed observationally in \citet{cleeves2015tw}. Paper I found that  the chemical reactions that lead towards deuterium-enriched H$_2$O proceed less efficiently without CR ionization. Thus, a significant fraction of the water in the solar system was likely inherited from the interstellar medium.

In the present paper, we extend this exploration to study the formation history of solar system organics with the \doh\ fractionation chemistry as our tool.  As can be seen in Figure~\ref{fig:dhplot}, both solar system and interstellar organics tend to have much higher \doh\ ratios than water, and have a much wider spread in values. While the term ``organics'' is less well-defined than is the case for water, we focus on the total hydrocarbon content in our chemical models, mostly in the form of simple organics with six atoms or fewer. We note that throughout the present paper, we will refer to organics and hydrocarbons interchangeably since we are focusing on the hydrogen-bearing organic species. We also follow the deuterium fractionation chemistry of two particular molecules, CH$_4$ and CH$_3$OH. These species are the most abundant simple organics in comets \citep{mumma2011} and have well-characterized ice abundances in the interstellar medium \citep[e.g.,][]{gibb2004,oberg2011spitz}. They furthermore trace comparatively different formation pathways. Collectively, the full set of simple organic species considered in our models may be potential precursors to macromolecular organics, and are more directly chemically linked to the important deuterium-fractionating reactions.  Similar to the approach taken in Paper~I, we simulate a scenario where the organic molecules start with the atomic \doh\ ratio (i.e., have no enrichment in \doh) to quantify to what extent disk chemistry contributes toward deuterium enrichment in solar system organics. In this regard, we take a different approach than previous works \citep[e.g.,][]{aikawa1999,aikawa2001,willacy2007,willacy2009} where more realistic interstellar \doh\ ratios are the initial conditions and the disk chemistry evolves forward \citep[see also][for an initially warm chemistry deuterium fractionation scenario]{aikawa1999}. The goal of this work is a simpler thought-experiment, i.e., what happens if nothing from the interstellar phase is preserved during star and disk formation, and disk chemistry alone is left to produce the solar system's organic content? Can disk chemistry reproduce the deuterium isotopic signatures in solar system organics? 

The paper is structured as follows. In Section~\ref{sec:methods}, we present our modeling methods, including our physical model with updates to the X-ray radiation transfer (Section~\ref{sec:physmod}) and to the fully deuterated reaction network used in our chemical calculations (Section~\ref{sec:chemmod}). We present our findings in Section~\ref{sec:chemresults}, and discuss these in relation to previous work along with their implications in Section~\ref{sec:discussion}. Finally, we summarize our results in Section~\ref{sec:summary}.

\section{Methods}\label{sec:methods}
\subsection{Physical Model}\label{sec:physmod}
We adopt the density and thermal structure model of the dust and gas that was used in Paper~I originally presented in \citet{cleeves2013a}.  Briefly, the model reflects a generic T~Tauri disk around a solar mass star ($T_{\rm eff}=4300$~K) with an X-ray luminosity of $L_{\rm XR}=10^{29.5}$~erg~s$^{-1}$. There are two dust size distributions, a large grain population with maximum size of 1~mm and concentrated near the midplane, and a second mixed population of 1 and 10 micron-sized grains distributed over the full gas scale height of the disk. Both grain populations have a minimum size of 0.005~$\mu$m. Based on the CR modeling of \citet{cleeves2013a}, and confirmed by \citet{cleeves2015tw}, we have adopted a relatively low CR ionization rate of $\zeta_{\rm CR} = 2\times10^{-19}$~s$^{-1}$ per H$_2$. This value coincides with the upper limit measured by \citet{cleeves2015tw}, although CR ionization rates in TW Hya may be substantially lower.  We have also included ionization by the decay of short-lived radionuclides (SLRs) at their initial abundances for the solar system, taking into account decay product losses. The SLR ionization rate has a maximum value of $\zeta_{\rm SLR} = 10^{-18}$~s$^{-1}$ in the inner disk, dropping to $\lesssim6-8\times10^{-19}$~s$^{-1}$ from $50-80$~AU where more decay products escape prior to ionizing the disk gas. The SLR ionization decreases over time with a net half-life of 1.2~Myr \citep{cleeves2014par}, such that at the end of the simulation the rate is about half of its initial value.
While we consider the time dependent chemical evolution, we consider the physical structure to be fixed and do not have turbulent mixing in the present models.

\begin{figure}
\begin{centering}
\includegraphics[width=0.47\textwidth]{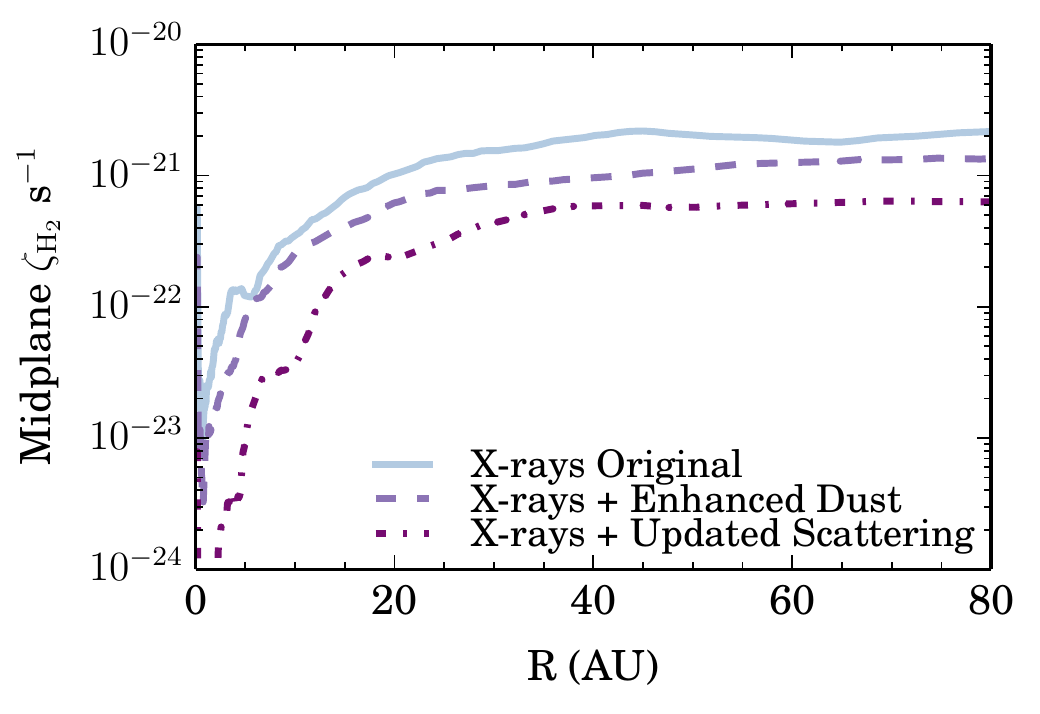}
\caption{Updated X-ray treatment from Paper~I.  The top (solid) curve shows the midplane ionization rate due to X-rays in the original model. The dashed curve shows the ionization rate when taking into account the enhanced dust mass in the midplane (in addition to the depleted dust mass in the surface, which was already considered). The dotted curve shows our updated treatment of X-ray scattering, where we had originally approximated the scattering as isotropic, but now include the proper directionality of the scattering angle. \label{fig:xray}}
\end{centering}
\end{figure}

The main update in the physical model is the treatment of the X-ray propagation, which is calculated using the code of \citet{bethell2011u} with the absorption cross sections of \citet{bethell2011x}.  In the original model in Paper~I, we took into account the spatial variations in the dust opacity to X-rays; however, we capped the opacity above a dust-to-gas mass ratio of 1:100, and thus underestimated the X-ray opacity in the midplane.  While X-rays are already highly attenuated upon reaching the midplane, by including the enhanced midplane dust the X-ray ionization rate per H$_2$ drops by a factor of $\sim2$, reducing the total midplane ionization (see Figure~\ref{fig:xray}, dash-dot line). 
We also improved the treatment of scattering in the X-ray radiation transport, where originally Thompson scattering was approximated as isotropic, it now more realistically includes preferential forward and backward scattering.  Taking this into account reduces the ionization in the midplane by a factor of $\sim1.2$. Altogether, the combined enhanced dust and improved scattering treatment reduce the X-ray ionization rate at the midplane by a factor of $\sim2.4$ lower than previously assumed (see Figure~\ref{fig:xray}).  It should noted, however, that both the original and updated X-ray values in the midplane are below that of the reduced CR rate assumed, and are not the dominant ionizing source in the coldest disk gas in these models.

\subsection{Chemical Model}\label{sec:chemmod}
To calculate the chemical abundances as a function of position and time, we use the chemical code of \citet{fogel2011} with updates as described in \citet{cleeves2014par}, including simple grain surface chemistry.  The original reaction network used in Paper~I included $\sim6200$ reactions and $\sim600$ species and was designed to predict the deuterium chemistry of water. For this work, we have expanded the network to include deuterium fractionation reactions for the carbon-bearing species, which greatly expands the size of the chemical network. We have substituted up to three H/D-atoms per molecule and deuterate molecules with three of fewer ``metals'' (i.e., non-hydrogen atoms). Statistical branching ratios for standard, non-fractionating reactions were assumed. The most important fractionation reactions for the conditions present in the bulk, cold disk are: 
\begin{equation}\label{eq:h2dp}
{\rm H_3^+ + HD \longleftrightarrow H_2D^+ + H_2 + \Delta E_1, }
\end{equation}
\begin{equation}\label{eq:ch2dp}
{\rm CH_3^+ + HD \longleftrightarrow CH_2D^+ + H_2 + \Delta E_2, }
\end{equation}
\begin{equation}\label{eq:dcop}
{\rm  HCO^+ + D \longleftrightarrow DCO^+ + H + \Delta E_3,}
\end{equation}
where all of the reactions are exothermic going from left to right.  The specific value of ${\rm \Delta E_1}$ depends on the spin of H$_3^+$ \citep{hugo2009} and is implemented as described in \citet{cleeves2014wat}.  For CH$_2$D$^+$, we take into account the ortho-to-para ratio of H$_2$ \citep{roueff2013}, assuming the H$_2$ o/p ratio is thermal \citep{lee2015,cleeves2014par}. We assume $\rm \Delta E_3=796$~K based on \citet{adams1985}. Beyond the reactions detailed above, there additional pathways that can lead to organic fractionation in the ices. In particular, \citet{hidaka2009} show that CH$_3$OH and H$_2$CO can become fractionated by abstraction (dissociation by reactions with H and D atoms), followed by D-atom addition. Because we are interested in the global \doh\ ratio across the total organic population and have a relatively simple ice chemistry, we do not include these abstraction-addition reactions in the present work, but they may play an important role in the fractionation between different isomers of CH$_3$OH, and provide additional pathways toward increasing \doh\ in CH$_3$OH and H$_2$CO in the presence of atomic D and H.
 
 The expanded network in total includes 15,711 reactions and 1,014 species. The initial abundances used in our model are originally based on \citet{fogel2011}, which were designed to reflect the molecular cloud chemical calculations from \citet{aikawa1999} as the starting point for the disk, and further updated as discussed in \citet{cleeves2014par}.  In the present work, we set the abundances of H$_2$O, CO$_2$, CH$_4$, and CH$_3$OH ices to reflect the measured {\em Spitzer} ice abundances presented in \citet{oberg2011spitz}, see Table~\ref{tab:chem}. For the initial \doh\ in organics and water, we have assumed the bulk value, or $\rm{\left[D/H\right]_{H_2}}=2.0\times10^{-5}$.  Both methane and methanol have gas phase and grain surface reaction pathways that are considered in the network and whose grain surface chemistry is dominated by hydrogenation of C and CO, respectively.  We include self-shielding of CO, H$_2$, HD and D$_2$ \citep[see][for details]{cleeves2014wat}.

\begin{deluxetable}{llll}
\tablecolumns{4} 
\tablewidth{0pt}
\tablecaption{Initial chemical abundances. \label{tab:chem}}
\tabletypesize{\footnotesize}
\tablehead{Species&Abundance&Species&Abundance}
\startdata
H$_2$ & $5.00\times10^{-1}$ & HD & $2.00\times10^{-5}$ \\
He & $1.40\times10^{-1}$ & H$_2$O(gr) & $2.40\times10^{-4}$ \\
HDO(gr) & $9.60\times10^{-9}$ & CH$_4$(gr) & $5.70\times10^{-6}$ \\
CH$_3$D(gr) & $4.56\times10^{-10}$ & CH$_3$OH(gr) & $1.80\times10^{-5}$ \\
CH$_3$OD(gr) & $1.44\times10^{-9}$ & N & $2.25\times10^{-5}$ \\
CN & $6.00\times10^{-8}$ & N$_2$ & $1.00\times10^{-6}$ \\
NH$_3$ & $8.00\times10^{-8}$ & HCN & $2.00\times10^{-8}$ \\
DCN & $4.00\times10^{-13}$ & H$_3$$^+$ & $1.00\times10^{-8}$ \\
H$_2$D$^+$ & $1.30\times10^{-10}$ & HD$_2$$^+$ & $1.00\times10^{-10}$ \\
D$_3$$^+$ & $2.00\times10^{-10}$ & CS & $4.00\times10^{-9}$ \\
SO & $5.00\times10^{-9}$ & Si$^+$ & $1.00\times10^{-9}$ \\
Mg$^+$ & $1.00\times10^{-9}$ & Fe$^+$ & $1.00\times10^{-9}$ \\
C$^+$ & $1.00\times10^{-9}$ & CO & $4.33\times10^{-5}$ \\
CO$_2$(gr) & $3.30\times10^{-5}$ & HCO$^+$ & $9.00\times10^{-9}$ \\
C$_2$H & $8.00\times10^{-9}$ & & 
\enddata
\end{deluxetable}

\section{Chemical Results} \label{sec:chemresults}
\subsection{ Organic \doh\  }

\begin{figure*}[ht!]
\begin{centering}
\includegraphics[width=0.95\textwidth]{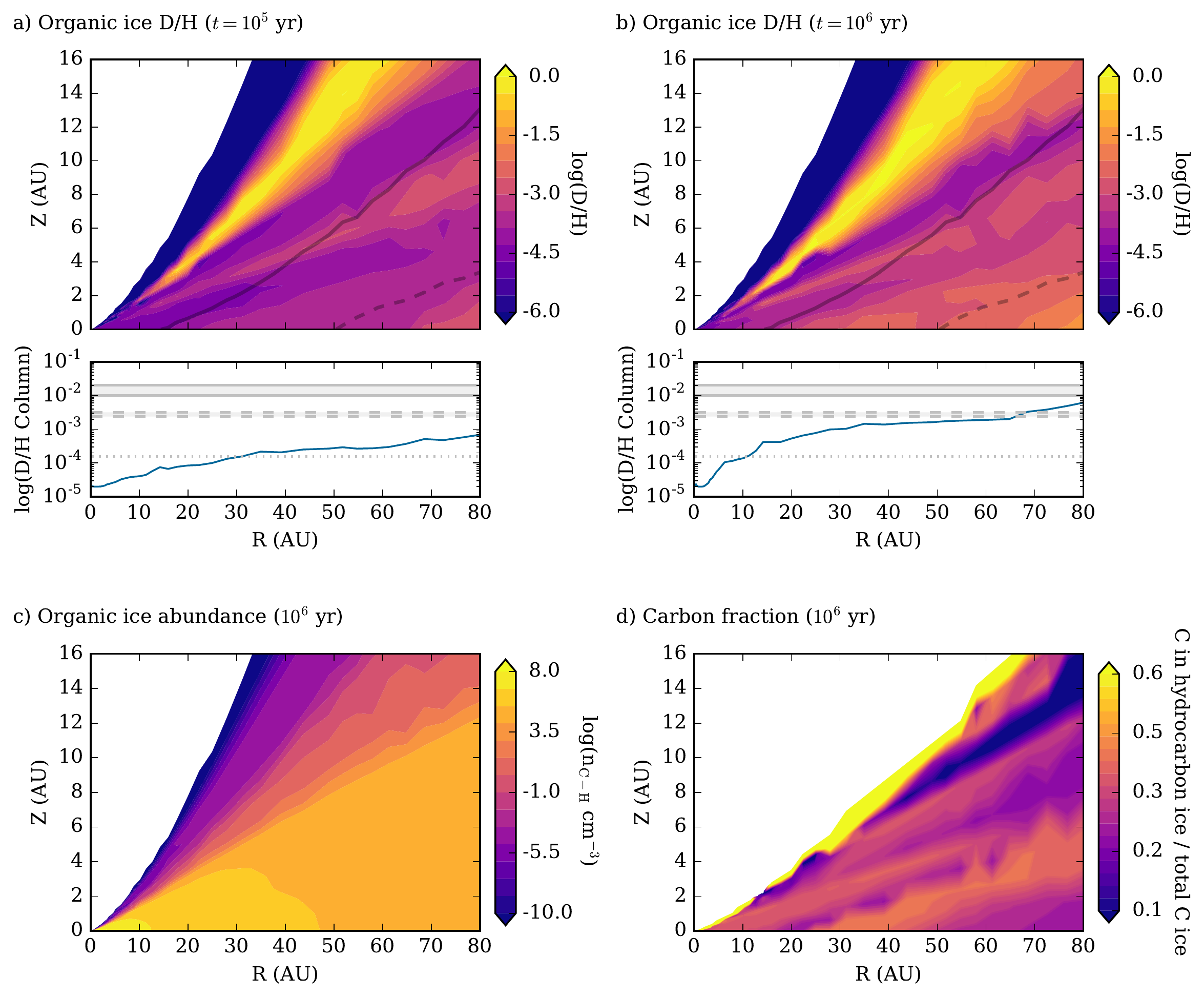}
\caption{Top panels: D/H across all hydrocarbon ice spatially ($R,Z$) and in vertically integrated column density for two time steps, a) $10^5$ years and b) $10^6$ years. In the spatial \doh\ plots, the solid (dashed) line shows the 30~K (20~K) temperature contour.  The grey lines in the vertically integrated plots provide points of comparison, where \doh\ in VSMOW is the dotted line, the range for meteoritic ``hot spots'' \citep{busemann2006} is between the dashed lines, and the inferred range for Orgueil radical compositions \citep{gourier2008} are given by the solid lines. As can be seen, no disk chemical model is able to reach \doh\ as measured in the Orgueil radicals, though organics are clearly enhanced above VSMOW \doh\ values at both early and late times in the outer disk. Panel c) shows the spatial number density of carbon in organic ice as a function of position, and panel d) shows the fraction of carbon in hydrocarbon (organic) ice compared to the total carbon ice budget.\label{fig:dhcarbfrac}}
\end{centering}
\end{figure*}

Figure~\ref{fig:dhcarbfrac} presents the main result of this work, where we calculate \doh\ for all hydrocarbons on grain surfaces, i.e, the global solid organic \doh.  In this plot, we show the total number of D-atoms that are associated with one or more C-atoms, compared to the same calculation for H-atoms. This calculation includes multiply-deuterated molecules. We show both the spatial distribution of \doh\ (top row of Fig.~\ref{fig:dhcarbfrac}a and b) and \doh\ derived from the vertically integrated column density (bottom row of Fig.~\ref{fig:dhcarbfrac}a and b) at chemical simulation times of a) $10^5$ and b) $10^6$ years.  The organic fractionation chemistry in the molecular layer ($z/r\sim0.1$) and closer to the midplane is dominated by the CH$_2$D$^+$/CH$_3^+$ reaction pathway, Eq.~(\ref{eq:ch2dp}). The organics present in the warm molecular layer have higher \doh\ than that found for water in Paper~I, highlighting the importance of Eq.~(\ref{eq:ch2dp}) at facilitating deuterium fractionation in the carbon reservoir even in warm ($\sim50-100$~K) gas \citep{favre2015}. This reaction, combined with the higher X-ray flux closer to the surface, results in more efficient organic deuteration in these layers compared to water. Even higher up in the disk, where $z/r\sim0.25$, deuterium fractionation is driven by reactions involving DCO$^+$/HCO$^+$ in the presence of H and D atoms, see Eq.~(\ref{eq:dcop}) and \citet{oberg2015d}.  In the surface layer, the difference in self-shielding efficiency between H$_2$ and HD leads to a higher photodissociation rate of HD in a layer where H$_2$ is able to self-shield, driving the gas phase atomic \doh\ ratio toward unity. However, the organic density in this layer is extremely low (see Figure~\ref{fig:dhcarbfrac}c).

Even though these additional pathways result in very high levels of \doh\ in organic species in the warm layers and the upper atmosphere, it should be noted that the midplane maintains relatively low \doh. The high mass density in the midplane compared to that above $z/r \gtrsim 0.1$ conspires such that the total \doh\ measured from the column densities more closely reflects the midplane values over the surface. However, conversely, these findings provide a cautionary note in interpreting observed \doh\ in species that trace gas closer to the surface since they probably do not reflect the bulk disk \doh. 

We also find that a significant fraction of the total carbon budget in the ices is in the form of organics. Figure~\ref{fig:dhcarbfrac}d shows the fraction of total carbon ice that exists as hydrocarbon ice in our models at 1~Myr, not including carbon grains or PAHs in this accounting. A substantial fraction, $\sim30-50\%$, of this ``volatile'' carbon-ice inside of 40~AU is in hydrocarbon form, while the rest is in CO and CO$_2$ ice depending on radial location. The part of the plot near the upper edge ($z/r\gtrsim0.2$) with very high hydrocarbon fraction (yellow) is in the form of relatively large molecular carriers, such as CH$_3$OH and CH$_2$CN ice.  Below this layer, CO$_2$ can readily form in the gas and freeze out resulting in a very low hydrocarbon ice fraction near $z/r\sim0.16$. The striped high ratios below this layer occur in regions where certain hydrocarbon species are favored, such as HCN, H$_2$CN, CH$_3$OH, HC$_3$N, and CH$_4$. Deeper in the cooler parts of the disk, CO and CO$_2$ ice are the main carbon-bearing ice constituents ($\sim75-80\%$), especially beyond the 40~AU in the midplane. 

\subsection{Chemical Dependence on Ionization Properties}

Because of the strong dependence on the efficiency of deuterium fractionation with the degree of ionization, we also present a series of chemical models that vary the CR and X-ray flux.  We have examined models with:
\begin{enumerate}
\item A reduced CR ionization rate or $\zeta_{\rm CR}\sim2.2\times10^{-19}$ s$^{-1}$ \citep[model SSX;][]{cleeves2013a} and a typical X-ray luminosity, $L_{\rm XR}=10^{29.5}$~erg~s$^{-1}$ (the standard model).
\item A further reduced CR rate, $\zeta_{\rm CR}\sim3.8\times10^{-22}$ s$^{-1}$ \citep[model TTX;][]{cleeves2013a}, with the same X-ray luminosity as (1). SLRs dominate but decay over time.
\item  An SSX CR rate with an order of magnitude higher X-ray luminosity, or $L_{\rm XR}=10^{30.5}$~erg~s$^{-1}$. 
\end{enumerate}
All of these models contain a time-decaying baseline ionization level due to SLRs calculated using the approach of \citet{cleeves2013b} with ``solar nebula-like'' abundances of SLR species as discussed in Section~\ref{sec:physmod}. 
The corresponding total midplane ionization rate at 1~Myr per H$_2$ at  $R=70$~AU for each of these models is 1) $6.14 \times 10^{-19}$~s$^{-1}$, 2) $3.97\times 10^{-19}$~s$^{-1}$, and 3) $6.22\times10^{-19}$~s$^{-1}$.

\begin{figure*}[ht!]
\begin{centering}
\includegraphics[width=1.0\textwidth]{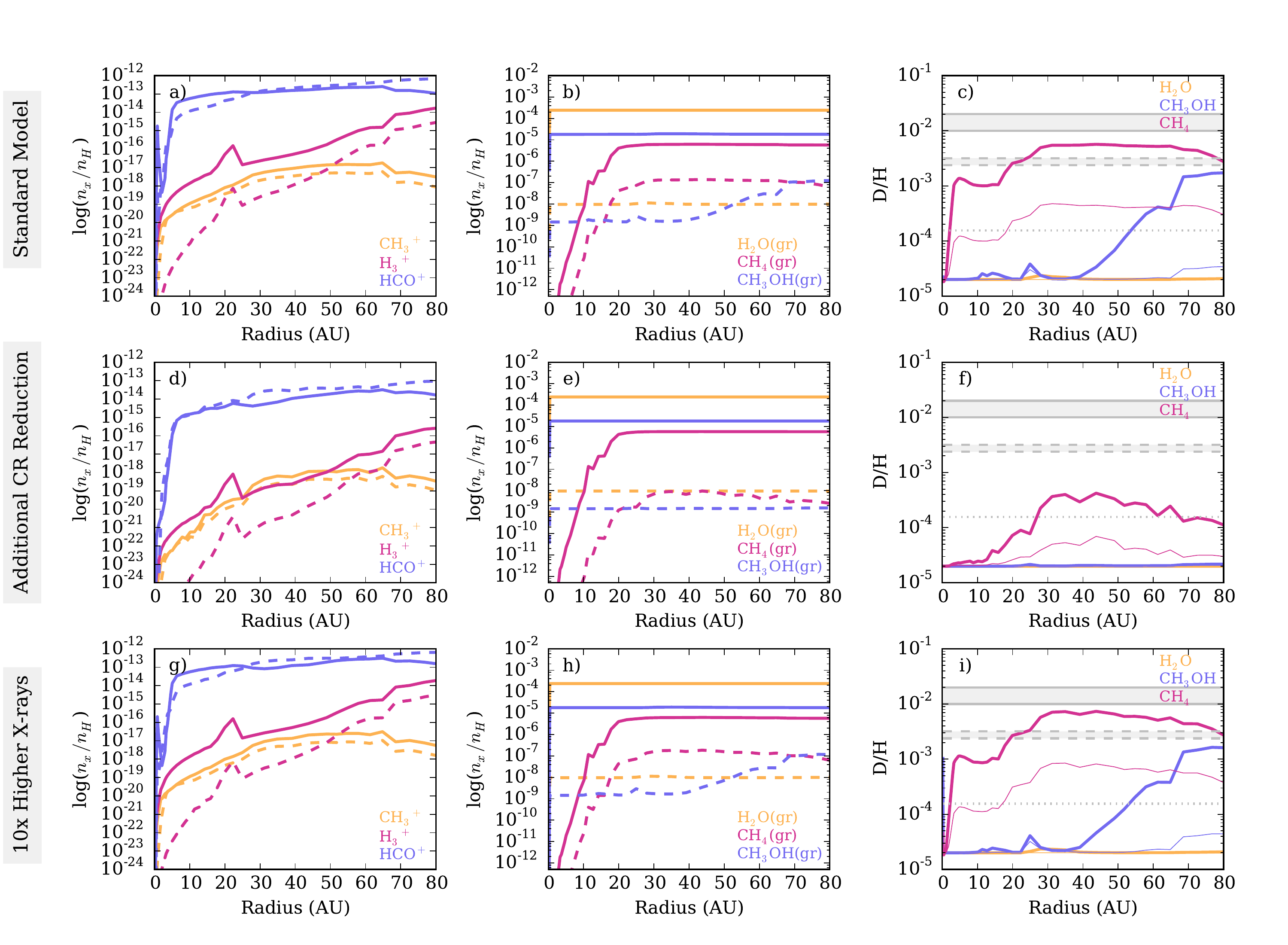} 
\caption{$\rm{\left[D/H\right]}$ in ions (precursors) and in grain-surface species  in the midplane. In the left two columns, the solid line shows the abundance of the main isotopologue and the dashed line is for the singly deuterated isotopologue. The rightmost column shows the \doh\ ratio in the ices, where the thick line now shoes \doh\ at $10^6$ years and the thin line is at $10^5$ years. In regions where the thin line is not visible, it overlaps the later stage, i.e., there has been no subsequent chemical evolution between these time steps.  Top row is a SSX CR model. Middle row is further reduced CR rate \citep[TTX; $\zeta_{\rm CR}\sim3\times10^{-22}$ s$^{-1}$;][]{cleeves2013a} with the same SLR and X-ray rate.  Bottom row is the SSX CR rate with a factor of 10 higher X-ray luminosity. Grey lines are the same as in Figure~\ref{fig:dhcarbfrac}. \label{fig:dhmidplane}}
\end{centering}
\end{figure*}

To examine the drivers of the deuteration chemistry, we examine in detail \doh\ of CH$_4$ and CH$_3$OH ice. By comparing these two species, we can study how the gas phase and grain surface pathways factor into the isotopologue ratios.  In particular, we consider the ratios of CH$_3$D/CH$_4$, CH$_3$OD/CH$_3$OH, and HDO/H$_2$O. We do not consider isomers in the current network, and these ratios are proxies for the amount of additional deuterium transferred to each of these species.  The calculated \doh\ ratios are shown in the rightmost column of Figure~\ref{fig:dhmidplane}.  Broadly, we find that both CH$_4$ and CH$_3$OH are more readily chemically fractionated compared to H$_2$O in all of our models, even at the midplane (see Figure~\ref{fig:dhmidplane}). However, the fractionation pathways are nonetheless curtailed by the low ionization environment of the disk. The ions, CH$_3^+$, H$_3^+$, and HCO$^+$ become extremely deuterium enriched approaching a \doh\ of unity (see Figure~\ref{fig:dhmidplane}, left column); however, their low absolute abundances inhibit their \doh\ ratios from being efficiently passed on to either of the products via gas phase or grain surface chemistry within Myr-timescales (Figure~\ref{fig:dhmidplane}, right column), similar to what was found for H$_2$O ice in Paper~I.  We also find that the \doh\ value depends strongly on the particular ionization rate and the chemical time. Time is an important consideration as the planet-formation process likely spans a considerable range \citep[the parent bodies of iron meteorites may have formed 0.1-0.6~Myr after CAIs and CI, CM, CR chondrites accreted over $\sim4$~Myr after CAIs;][]{kruijer2014}. 

\begin{figure}
\begin{centering}
\includegraphics[width=0.48\textwidth]{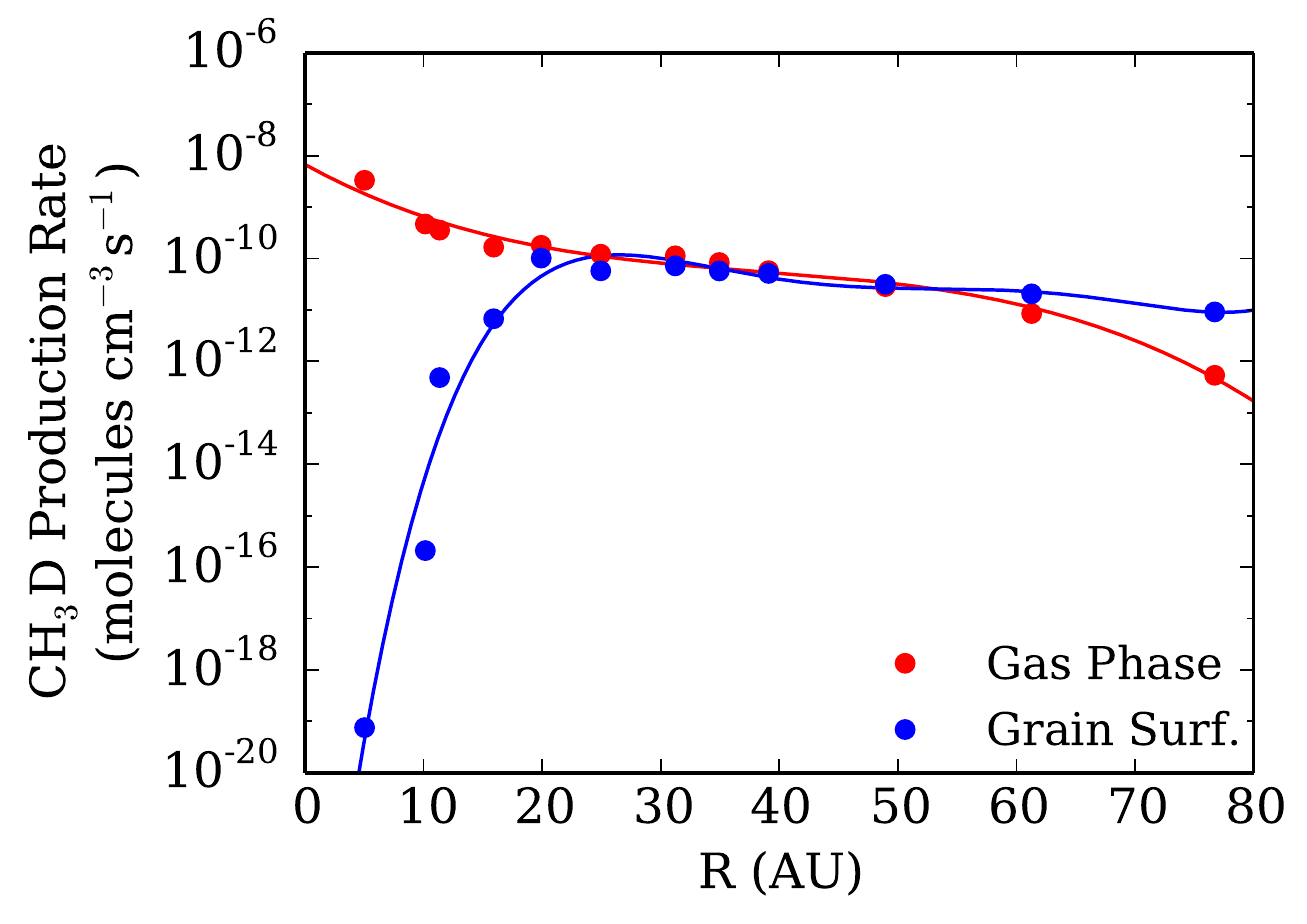}\caption{Singly deuterated CH$_4$, CH$_3$D, formation by gas phase (red) and grain surface (blue) channels in the disk midplane versus radial distance from the star. Rates are calculated at the points, while the lines help guide the eye. \label{fig:calcrate}}
\end{centering}
\end{figure}

In our standard model (top row of Figure~\ref{fig:dhmidplane}), CH$_4$ ice becomes abundant at $\gtrsim15-20$~AU, and is deuterium-enriched in the midplane throughout the disk \citep[the binding energy assumed is $E_b=1090$~K or a desorption temperature of $T_d\sim22$~K;][]{herrero2010}.  
Note that while the \doh\ ratio of CH$_4$ ice is elevated inside of $\sim20$~AU, the absolute amount of CH$_4$ ice at this location is low and thus does not contribute significantly to the overall solid organic \doh\ budget in the disk. We find that the dominant CH$_4$ formation channel transitions from being primarily gas phase inside of $R<20$~AU, to being a nearly equal combination of gas-phase and grain-surface production outside of this region (see Figure~\ref{fig:calcrate}). The \doh\ value of this molecule is also strongly impacted by time, where \doh\ increases by about an order of magnitude between $10^5$~years and $10^6$~years, peaking at $5\times10^{-3}$ at late times for our standard model. 

CH$_3$OH, on the other hand, is a primarily grain-surface chemistry product for the bulk mass of the disk \citep[though some gas-phase routes are available closer to the X-ray irradiated disk surface, see][]{bergin2014far}.  CH$_3$OH becomes enriched only outside of $R=40$~AU and only at late times, $10^6$~years, even though it is abundant in ice from $\sim1$~AU outward (see middle column, Figure~\ref{fig:dhmidplane}).    The maximum \doh\ in CH$_3$OH attained in our models after $10^5$ years is $2\times10^{-5}$, while after $10^6$~years is $2\times10^{-3}$ beyond 70~AU, dropping steeply inward. 
While the present models do not include the abstraction/addition reactions for CH$_3$OH discussed in Section~\ref{sec:chemmod}, we see that the low ionization environment curtails the abundances of free atomic H and D, and thus we expect abstraction/addition to also be less efficient under low ionization conditions considered in this work.

Based on these models, it is clear that CH$_3$OH and CH$_4$ have very different \doh\ values and radial distributions when synthesized by disk chemistry  alone. We also confirm the results of Paper~I with this larger deuterated network, namely that $\rm{\left[D/H\right]_{H_2O}}$ remains unaltered by disk chemistry for the reduced CR-ionization models.

Further decreasing the CR ionization rate from  $\zeta_{\rm CR}\sim2\times10^{-19}$ s$^{-1}$ to $\zeta_{\rm CR}\sim3\times10^{-22}$ s$^{-1}$ (see Fig.~\ref{fig:dhmidplane}, middle row) fully inhibits the CH$_3$OH deuterium enrichment, and significantly reduces that of CH$_4$.  Under the lower CR ionization conditions, the maximum value of $\rm{\left[D/H\right]_{CH_4}}$ is $7\times10^{-5}$ ($4\times10^{-4}$) at $10^5$ ($10^6$) years.  The reason for the sharp decrease in deuterated CH$_4$ and CH$_3$OH at 1 Myr is due to the time-evolving SLR ionization rate. For Model 1, the production of the deuterated organic species sharply increases from 0.1 to 1 Myr by an order of magnitude. For Model 2, during this phase the ionization (which is dominated by SLRs) begins to shut off due to time-decay, curtailing this phase of late organic formation in Model 2.  If we alternatively use our standard SSX CR rate but increase the X-ray luminosity by an order of magnitude (Fig.~\ref{fig:dhmidplane}, bottom row), \doh\ in CH$_4$ increases, but only by a factor of $\sim1.4$. As is clear from the midplane ionization values discussed above, CRs along with SLRs, appear to be the dominant midplane ionizing agents that set the degree of deuterium fractionation in these simple organics, but the efficiency of this process is still sensitive to the absolute value of the combined ionization rate, whether it is  $6\times10^{-19}$ s$^{-1}$ versus a factor of two lower, which is not the case for water where cold water formation is shut off at higher total ionization rates.

\begin{figure*}
\begin{centering}
\includegraphics[width=0.7\textwidth]{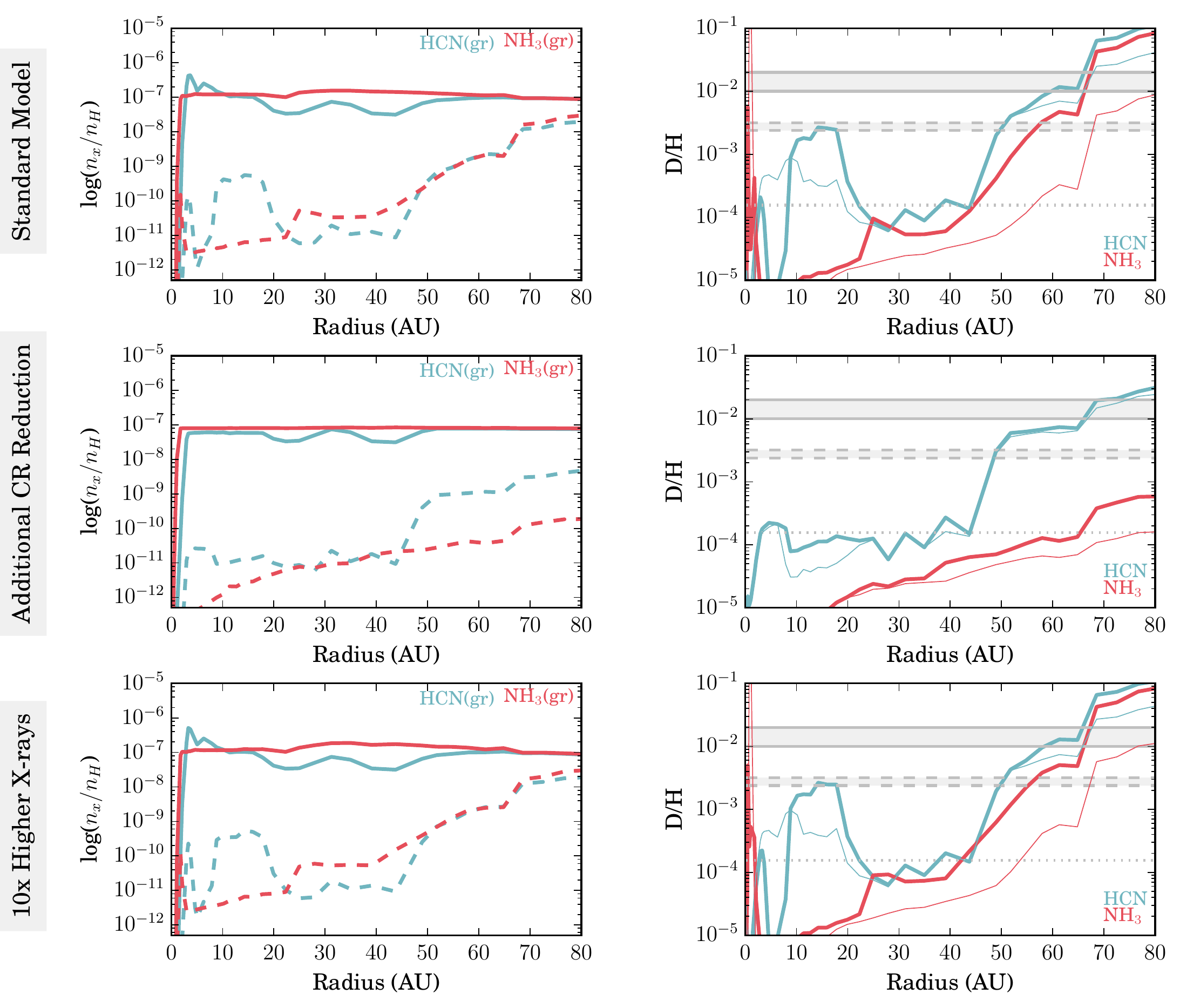}
\caption{Midplane \doh\ in hydrogenated nitrogen-bearing molecular ices.  Figure is labeled as in Figure~\ref{fig:dhmidplane}, where the solid line represents the abundance of the main isotopologue and the dashed line is the singly deuterated isotopologue. In the right panel the D/H ratios are shown for two time steps, $10^5$ years (thin line) and $10^6$ years (thick line). Solar system comparison values are shown as the horizontal gray lines as in Figure~\ref{fig:dhmidplane}.  \label{fig:dhnit}}
\end{centering}
\end{figure*}

Finally, we also consider the deuterium-enrichment of nitrogen-bearing molecules, specifically NH$_3$ and HCN (see Figure~\ref{fig:dhnit}).  The nitrogen chemistry parallels that of CH$_4$ and CH$_3$OH, where the HCN is formed mainly by gas-phase chemistry while NH$_3$ is formed on grains. The fractionation chemistry via gas-phase channels is more efficient than that for hydrogenation on grain-surfaces in our models, and therefore for both the carbon and nitrogen-reservoirs, the molecules with dominant gas phase channels are more readily D-enriched.

\section{Discussion}\label{sec:discussion}

\subsection{Disk Deuterium Fractionation Pathways}

We have identified two major factors that lead to the higher degree of deuterium enrichment in protoplanetary disk organics as compared to H$_2$O.
The first is the higher volatility and availability of the main carbon reservoir in the disk, CO, compared to atomic oxygen (the main H$_2$O precursor for grain-surface formation).  CO photodissociation liberates some fraction of the carbon, allowing it to react in the gas phase to form hydrocarbons (C$\rm _X$H$\rm{_Y}$), including CH$_4$, and to react on grains through C-hydrogenation \citep[e.g.,][]{bergin2014far,furuya2014,reboussin2015}. CH$_3$OH is formed from hydrogenation of CO ice, which is a major component of the disk midplane ices especially beyond 20~AU. The relatively high abundance of CO ice compared to atomic oxygen ice is what enables CH$_3$OH to form more efficiently than H$_2$O with a fractionated \doh\ ratio. 

\begin{figure*}
\begin{centering}
\includegraphics[width=0.65\textwidth]{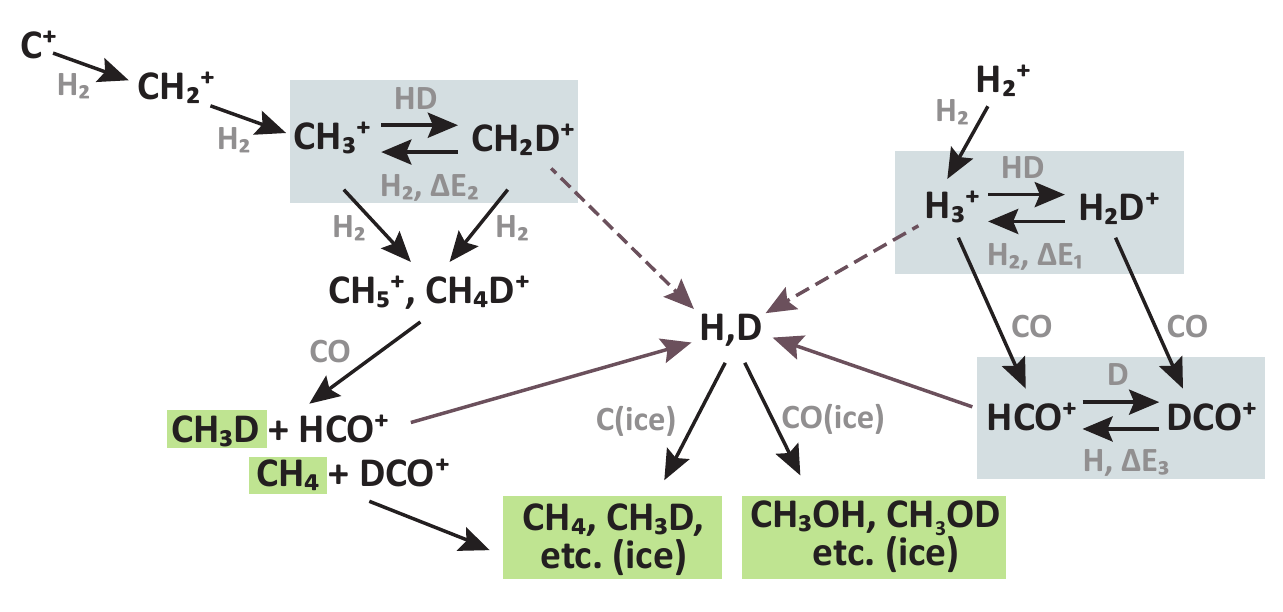}
\caption{Chemical pathways that lead to deuterium fractionation for carbon-bearing species in the gas and on grains.  Brown arrows indicate recombination pathways that produce H and D-atoms.  The dashed brown lines show the main reactions resulting in  deuterium fractionation in atoms deeper in the disk, while the solid brown lines highlight the contribution from DCO$^+$ and HCO$^+$ recombination, which are an important source of deuterium fractionation in atoms closer to the surface where there is both CO in the gas and selective self-shielding of H$_2$ compared to HD. Only the main reaction products important for this paper are pictured here for clarity.   \label{fig:ch2dp}}
\end{centering}
\end{figure*}

The second factor leading to higher organic \doh\ is even more fundamental, namely the deuterium-fractionation chemistry is generally more favorable for organics than for water. This point applies across astrophysical environments, both in the disk and the molecular cloud core. The higher exothermicity of Eq.~(\ref{eq:ch2dp}) naturally produces higher \doh\ in CH$_3^+$ than in H$_3^+$ at $T>10$~K, such that organics (which can form via chemistry involving both CH$_3^+$ and H$_3^+$) are able to attain generally higher \doh\ than H$_2$O \citep{aikawa1999}. This chemical feature is a potential explanation for the relatively higher \doh\ values in solar system organics than water shown in Figure~\ref{fig:dhplot}. This point is especially important in the disk, where now more of the disk volume can participate in deuterium fractionation of organics, including the warm ($T\sim50-100$~K), X-ray irradiated disk surface. This point also touches on the fact that there are simply more fractionation pathways impacting \doh\ in organics as is illustrated in Figure~\ref{fig:ch2dp}. As a result, we expect more variation in organic \doh\ based on a specific molecule's formation reactions, which is perhaps reflected in the relatively higher spread in organics compared to H$_2$O (Figure~\ref{fig:dhplot}). The impact of these additional gas-phase channels can be seen in the models based on much greater levels of \doh\ attained by CH$_4$ in the inner disk ($\lesssim40$~AU) as compared to CH$_3$OH.  

Another interesting facet of the chemistry is the importance of radiative association, i.e., a two body reaction that results in a single product and the emission of a photon. The rates of radiative association are enhanced at the densities relevant for the molecular layers of disks, i.e., $n_{\rm H_2}\sim10^{8}-10^{10}$~cm$^{-3}$, and it is one of the most important channels for the formation of CH$_4$ and CH$_3$D, initiated with: 
\be{\rm CH_3^+ + H_2 \rightarrow  CH_5^+ + }~h\nu\ee
and
\be{\rm CH_2D^+ + H_2 \rightarrow  CH_4D^+ + }~h\nu.\ee
After formation, the CH$_3^+$ and CH$_4$D$^+$ products recombine with electrons, charged grains, or CO, forming CH$_4$ and CH$_3$D, along with DCO$^+$ and HCO$^+$, e.g.,:
\be{\rm CH_4D^+ + CO \rightarrow  CH_3D + HCO^+},\ee
\be{\rm CH_3D_2^+ + CO \rightarrow  CH_3D + DCO^+},\ee 
\be{\rm CH_4D^+ + electron \rightarrow  CH_3D + H,}\ee 
and
\be{\rm CH_4D^+ + grain^- \rightarrow  CH_3D + H}.\ee
These pathways directly link the deuterium fractionation of CH$_2$D$^+$ to that of newly produced CH$_3$D. Yet even so, their absolute contribution to the disk-produced CH$_3$D/CH$_4$ is hampered by low CR ionization rates.  Consequently, the \doh\ in CH$_4$ does not reach the \doh\ values in CH$_3^+$ during the disk lifetime (see Figure~\ref{fig:dhmidplane}). 

There is an additional mechanism in the literature that has been proposed for the direct deuterium fractionation of macromolecular organics (here noted by ${\rm X{-}CH}$ and ${\rm X{-}CD}$) by deuterium transfer from the H$_2$D$^+$ ion, namely: 
\be{\rm H_2D^+ + X{-}CH \rightarrow  H_3^+ + X{-}CD } \label{chpath} \ee
\citep[e.g.,][]{remusat2006}. \citet{robert2011} explored this reaction pathway experimentally by exposing an organic covered substrate to D$_{\rm X}^+$ ions.  While we do not have macromolecular organics in our model to test this directly, we emphasize the overall abundances of H$_2$D$^+$ and H$_3^+$ in the cold, dense midplane of the disk (as well as CH$_2$D$^+$ and CH$_3^+$ ions, for that matter) are very low, and we do not expect this reaction to be efficient for the conditions in our model.  Furthermore, whether or not Eq.~(\ref{chpath}) proceeds efficiently will be very sensitive to the charge of the target molecule/substrate. In the instance that the $\rm X{-}CH$ is part of an organic coating on a grain, the grain itself will become rapidly negatively charged under normal conditions in the disk. Consequently, a more likely reaction than Eq.~(\ref{chpath}) would be dissociative recombination of the parent ion, i.e.,:
\be{\rm H_2D^+ + X{-}CH^- \rightarrow  HD + H + X{-}CH } \label{chpath1} \ee
or
\be{\rm H_2D^+ + X{-}CH^- \rightarrow  H_2 + D + X{-}CH } \label{chpath2} \ee
such that no deuterium is transferred to the $\rm X{-}CH$.  Even in the case where the organic is not in the solid state, one may draw an analog with gas-phase PAH chemistry, where PAHs also become rapidly negatively charged, and correspondingly will also dissociate the parent ion upon interaction \citep[e.g.,][]{wakelam2008}.  This process is fully analogous to grain-surface recombination with charged grains \citep[e.g.,][]{aikawa1998,aikawa1999}.

Even though our models do not include more complex forms of carbon (refractory carbon grains, PAHs, etc.), we can speculate regarding their formation origins as traced by \doh\ based upon our results for the simple organics. The main reactions in the literature for deuterium fractionation of PAHs or other carbon-based macromolecular compounds include reactions with ions as discussed above  \citep[e.g.,][]{tielens1997}, grain-surface hydrogenation from a deuterium-rich gas reservoir \citep{charnley1997}, and UV irradiation in an ice mixture with other deuterium-enhanced molecules, such as H$_2$O \citep{sandford2000}. The first two reactions mirror those affecting deuterium fractionation in simple organics, and therefore we would expect these reactions to be similarly inefficient in the low-ionization disk environment. The third requires UV photons, and so its efficiency would depend on the turbulent cycling time of ice-coated grains into the surface layers, whose role is still uncertain \citep[see recent limits and discussion of][]{flaherty2015}.  Correspondingly, both the formation of complex, macromolecular organics from simpler organics along with their \doh\ ratios may be difficult to explain by non-turbulent disk chemistry alone, though further exploration is necessary.

We emphasize that the specific value of \doh\ reached in our models is very sensitive to the ionization model used.  We have used an observationally motivated low CR ionization rate, where the measured rate set an upper limit to the CR rate of $\zeta_{\rm CR}\lesssim2\times10^{-19}$~s$^{-1}$ per H$_2$ \citep{cleeves2015tw}. In reality, the CR rate may be far lower.  As can be seen in Figure~\ref{fig:dhmidplane}, middle row, further reducing the total ionization rate by less than a factor of two resulted in a lower \doh\ in CH$_4$ by over an order of magnitude, and fully curtailed the deuterium fractionation of CH$_3$OH.  It may thus be the case that even moderate deuterium enrichments in organics \citep[like those present in meteoritic hot-spots;][]{busemann2006} are difficult to achieve in disks even beyond 20~AU. More sensitive observations of ions in protoplanetary disks, including H$_2$D$^+$ and N$_2$D$^+$ \citep{cleeves2014par}, will be central to constraining exactly how much ionization-driven disk chemistry is able to facilitate the deuterium fractionation of simple organics.

\subsection{Towards Understanding the Origins of Solar System Organics}

While Figures~\ref{fig:dhcarbfrac} and \ref{fig:dhmidplane} demonstrate that organics are more readily isotopically fractionated than H$_2$O, it is important to note that all of the ionization models we consider are unable to chemically reproduce the most highly deuterium enriched species in the solar system, the Orgueil radicals, whose \doh\ ratios have been estimated to be $1.5\pm0.5 \times 10^{-2}$.  Likewise, our standard model is unable to reproduce the \doh\ values in meteoritic hot-spots reported in \citet{busemann2006}, with more stringent limits provided by better future disk ionization models. On the other hand, high degrees of deuterium enrichment in organics are easily obtained in the dense ISM (see Figure~\ref{fig:dhplot}) where temperatures are cold and the molecular gas is exposed to high levels of galactic CRs. For example, \citet{roberts2002} measured \doh\ in HCN (in the gas via ice sublimation) toward protostellar cores with values ranging $\sim(2-6)\times10^{-2}$ and H$_2$CO with \doh\ ratios of $\sim(3-7)\times10^{-2}$, far higher than \doh\ in cometary organics. This highly deuterium-enriched interstellar material could thus be the starting point for the disk, followed by chemical processing down to lower \doh\ values, perhaps by transport into/from warm regions during the disk's formation \citep{aikawa1999,yang2013}, later (disk) stage mixing \citep{furuya2013,albertsson2014}, or heating during accretion outbursts \citep{owen2015}. Regardless of the processing mechanism(s), the low deuterium fractionation efficiency in our disk models require at least some inheritance of deuterium-enriched interstellar organic material in the protoplanetary disk. In the context of Paper~I, we found that a substantial inheritance of interstellar H$_2$O was required to explain solar system \doh\ ratios. Complex organics have similar if not lower volatility and thus, if some H$_2$O survived the solar system's formation, it is not surprising that a similar fraction of organics could have survived as well. 

We also find that the greater number of channels that lead to deuterium fractionation in the gas result in very different \doh\ values depending on the particular organic species under consideration. There are also additional chemical channels available in disks, e.g., radiative association, due to relatively high densities present compared to the ISM. Thus, studying relative differences in \doh\ in different simple organics in comets, such as CH$_4$, HCN, H$_2$CO, etc., may help shed light on their formation environments, individually and collectively. 

Comets, given their nearly pristine record of the early solar system, provide interesting constraints on our models. There are few existing measurements of organic \doh\ in comets to date (see Figure~\ref{fig:dhplot}); however, Hale-Bopp, an Oort cloud comet, has a measurement of \doh\ in HCN with a value of $(2.3\pm0.4)\times10^{-3}$ \citep{eberhardt1995}.  Oort cloud comets are thought to form in the modern-day Uranus-Neptune region of our solar nebular protoplanetary disk \citep[see review of][]{dones2004}, i.e., $\sim 20-40$~AU. In this region, all of our models find that the disk produces a \doh\ of $\sim10^{-4}$ in HCN in 1~Myr, more than an order of magnitude below the measured cometary value. At face value, the data would thus suggest some degree of interstellar inheritance of deuterated HCN from the parent molecular cloud later incorporated into comets. If we take the \doh\ measured in HCN from the \citet{roberts2002} work, $\sim2\times10^{-2}$ and $\sim10^{-4}$ for the disk \doh, Hale-Bopp would require at least $>10\%$ interstellar HCN. This value is higher if Hale-Bopp formed earlier ($<1$~Myr) or if our primordial cloud had lower DCN/HCN. This fraction would also be higher for other molecular species primarily formed in the ice where the fractionation chemistry is less efficient (however, see the note about the abstraction-addition pathways in Section~\ref{sec:chemmod}).

However, in detail, the exact amount of inheritance will depend on many factors, especially the timescales for radial drift compared to the timescales for cometesimal formation and the relevant chemical timescales. A more stringent limit would be provided by \doh\ measurements of CH$_3$OH in comets, where we expect no disk-derived deuterium enrichment inside of 40~AU even at late (1~Myr) times, with no substantial enrichment at any location in the disk even at $10^5$~years. While dependent on the particular assumptions, recent models find that radial drift happens rapidly, such that in $\lesssim10^5$~years, the radial dust distribution is compacted by almost a factor of $\sim2$ \citep{birnstiel2014}. Such short timescales combined with our model results would imply that any deuterium enrichment in CH$_3$OH present in comets would reveal interstellar inheritance of organic material.

\subsection{Comparison to Previous Work}

In this work, we have explored deuterium chemistry in a new, observationally motivated disk ionization model starting off with the molecular \doh\ set to be the elemental \doh\ value (no initial deuterium enrichment) as a simple thought experiment. There have been a number of related previous works on deuterium chemistry in disks, where the initial conditions are more realistic, i.e., are typically set from a molecular cloud model at $\sim10$~K. Consequently, in these models the initial \doh\ ratio in the model starts out high and is subsequently evolved forward in time. While the results between this work and previous works will be different in an absolute sense given the differences in initial conditions, we can compare the relative chemical trends from the models. 

\citet{aikawa1999} first looked at deuterium chemical evolution as disk matter migrates inward at the viscous accretion rate in the midplane. They adopted the parameters of a typical T~Tauri disk with a standard ``dense gas'' CR ionization rate of $1.3\times10^{-17}$~s$^{-1}$ per H$_2$. In this work, the authors recognized the importance of the higher exothermicity of the CH$_2$D$^+$ fractionation reaction, Eq.~(\ref{eq:ch2dp}), compared to the H$_2$D$^+$ reaction, Eq.~(\ref{eq:h2dp}) in the deuterium chemistry of CH$_4$, even with the lower previous measurement for the CH$_2$D$^+$ back reaction energy, $\Delta E_2$ in Eq~(\ref{eq:ch2dp}). In addition to the radially evolving (accreting) models, they ran single condition (temperature, density) to isolate these effects, where chemistry at 30~K still fractionates HCN and CH$_4$, but not water. With the updated back reaction energy used in the present work, we find similar results even at temperatures exceeding 50~K. \citet{aikawa1999} also first explored a lower CR ionization rate ($1.3\times10^{-18}$~s$^{-1}$) to reduce the \doh\ ratio in H$_2$O to better match with comets (a factor of two lower \doh\ than the standard CR model), illustrating the importance of disk ionization for regulating deuterium chemistry. Likewise, they explored chemical models of hot material (with low \doh) flowing outward, and find that ion-neutral chemistry after $10^6$ can reach \doh\ in H$_2$O as high as $2\times10^{-4}$, and higher for organics, highlighting the importance of these chemical pathways. In \citet{aikawa2001}, the authors moved from the 1D accreting parcel model of \citet{aikawa1999} to a full 2D static disk model. They include X-ray ionization at a relatively high level, $L_{\rm XR}=10^{31}$~erg~s$^{-1}$, a factor of 30 higher than the present work, along with a standard cosmic ray ionization rate. They find that the \doh\ ratios determined from the CH$_4$ and H$_2$CO (a methanol precursor) column densities do not decrease inwards also because of the CH$_2$D$^+$ pathway operating in the warm gas, consistent with our findings between $R=30-80$~AU, see Figure~\ref{fig:dhmidplane}, rightmost column.

In a pair of papers, \citet{willacy2007} and \citet{willacy2009} explore the deuterium chemistry in the outer and inner disk respectively, including molecule formation on grain surfaces and multiply deuterated molecules. They also include the ionization by decay of $^{26}$Al in both papers, and X-ray ionization in the latter, focusing on dense gas ionizing agents. 
These papers confirm the trend that the hydrocarbon bearing species are more strongly fractionated by disk chemistry (increasing in \doh\ by factors of $\sim2-3$ from their initial values in their ``C'' model, which contains all of the relevant physics), while for the same model, water only increases by a factor of 1.2 in the outer disk. We find less fractionation in water based on having lower ionization and less ``volatile'' oxygen in our models, but the trend towards higher organic fractionation is still the same. The paper also qualitatively explores variations in the ionziation rate indirectly by varying the disk mass, which leads to higher opacity to photons/particles and faster ion-recombination. There is also a similar  enhancement in hydrocarbons near the disk surface where the disk is UV irradiated that we see in Figure~\ref{fig:dhcarbfrac}d.

Finally, all of these models, including ours, do not include mixing \citep[though][does radially evolve gas inward]{aikawa1999}. Correspondingly, these results are relevant for a laminar or very low turbulence disk model. With mixing, deuterium fractionation of molecular species can be strongly enhanced by the delivery of parent molecules, such as oxygen, to the cold midplane \citep[e.g.,][for the case of water]{furuya2013,albertsson2013}, or decreased by mixing deuterated ices into warmer regions where they gradually lose their isotopic signature. We would expect that mixing could also enhance the ability for organics to become deuterium enriched by moving carbon into the cold midplane; however, we expect the effect to be less strong than for water because fractionation can proceed even at $\sim50$~K closer to the surface. Simultaneously, observations of turbulence in disks are placing ever more stringent upper limits on the amount of mixing present in the outer disk \citep{flaherty2015}, and as such it is still an open and important question to what extent turbulence impacts deuterium chemistry in disks for both water and organics.

 \section{Summary}\label{sec:summary}
We have explored the contribution from disk chemical processes toward deuterium enrichment of simple organic molecules using an updated treatment of disk ionization for a static disk model. We find that:
\begin{enumerate}
\item Simple organics (CH$_4$ and CH$_3$OH) are more efficiently deuterium-enriched by disk chemistry than H$_2$O. 
\item The efficiency of  deuterium fractionation of organics is highly sensitive to the specific molecule being studied, the ionization model used, and the specific chemical time in the simulation. Better constraints on disk ionization properties (along with radial drift timescales, planetesimal formation timescales and disk mixing efficiency) will provide stronger limits on the ability for disk chemistry to fractionate deuterium in pre-planetary organics.
\item The channels for gas-phase deuterium enrichment are more efficient than those on grain surfaces. The efficient gas-phase channels enable CH$_4$ to become more deuterium enriched than CH$_3$OH by disk chemistry.  As a consequence, both the amount of  deuterium fractionation and the radial distribution of \doh\ are very different for these two species.  We find similar results for N-bearing species, specifically HCN vs. NH$_3$. Thus, future measurements of \doh\ in different simple organics in comets compared to disk chemical models may help shed light on these species parent formation environments.
\item While both CH$_3$OH and H$_2$O are formed in similar ways, i.e., by hydrogenation of CO and O ices, CH$_3$OH is more easily deuterium fractionated than water. This behavior is because of the high abundance of CO ice present in the outer disk, where even a low flux of H and D atoms are able to form deuterium-enriched CH$_3$OH at the level of \doh\ of $\sim10^{-3}$ at late times. With relatively little atomic oxygen ice present in our models, water is not efficiently formed/re-formed and the \doh\ is not altered from the initial value in the disk midplane.
\item When considering the bulk hydrocarbon \doh, none of the ionization models considered in this work are able to reach the extreme levels of \doh\ inferred for the Orgueil radicals \citep{gourier2008}, and have trouble with the \doh\ values in meteoritic hot spots and cometary HCN inside of $\sim20$~AU. Further reducing the CR rate also prevents the disk from reaching \doh\ levels in meteoritic hot-spots at all disk radii.  Thus, some interstellar inheritance is required to explain the most extremely D-enriched organic materials present in solar system bodies. 
\end{enumerate}

\acknowledgements{Acknowledgements: The authors are thoroughly grateful to the anonymous referee, whose comments and careful consideration significantly improved the manuscript. LIC and EAB acknowledge the support of NSF grant AST-1008800 and the Rackham Predoctoral Fellowship. CA was partially supported by NASA Cosmochemistry grant NNX14AJ54G. FD and EAB acknowledge support of grant NNX12A193G from the NASA Origins of Solar Systems Program. TJH acknowledges support from STFC Consolidated Grant ST/M00127X/1. K.I.\"O. also acknowledges funding from the Simons Collaboration on the Origins of Life (SCOL). Support for the writing of this manuscript was provided by NASA through Hubble Fellowship grant HST-HF2-51356.001-A awarded by the Space Telescope Science Institute, which is operated by the Association of Universities for Research in Astronomy, Inc., for NASA, under contract NAS 5-26555. }


\begin{thebibliography}{68}

\expandafter\ifx\csname natexlab\endcsname\relax\def\natexlab#1{#1}\fi

\bibitem[{Adams \& Smith(1985)}]{adams1985}
Adams, N.~G. \& Smith, D. 1985, \apj, 294, L63

\bibitem[{{Aikawa} \& {Herbst}(1999)}]{aikawa1999}
{Aikawa}, Y. \& {Herbst}, E. 1999, \apj, 526, 314

\bibitem[{{Aikawa} \& {Herbst}(2001)}]{aikawa2001}
---. 2001, \aap, 371, 1107

\bibitem[{{Aikawa} {et~al.}(1998){Aikawa}, {Umebayashi}, {Nakano}, \&
  {Miyama}}]{aikawa1998}
{Aikawa}, Y., {Umebayashi}, T., {Nakano}, T., \& {Miyama}, S. 1998, Faraday
  Discussions, 109, 281

\bibitem[{{Albertsson} {et~al.}(2014){Albertsson}, {Semenov}, \&
  {Henning}}]{albertsson2014}
{Albertsson}, T., {Semenov}, D., \& {Henning}, T. 2014, \apj, 784, 39

\bibitem[{{Albertsson} {et~al.}(2013){Albertsson}, {Semenov}, {Vasyunin},
  {Henning}, \& {Herbst}}]{albertsson2013}
{Albertsson}, T., {Semenov}, D.~A., {Vasyunin}, A.~I., {Henning}, T., \&
  {Herbst}, E. 2013, \apjs, 207, 27

\bibitem[{{Alexander} {et~al.}(2012){Alexander}, {Bowden}, {Fogel}, {Howard},
  {Herd}, \& {Nittler}}]{alexander2012}
{Alexander}, C.~M.~{\od}., {Bowden}, R., {Fogel}, M.~L., {Howard}, K.~T.,
  {Herd}, C.~D.~K., \& {Nittler}, L.~R. 2012, Science, 337, 721

\bibitem[{{Alexander} {et~al.}(2007){Alexander}, {Fogel}, {Yabuta}, \&
  {Cody}}]{alexander2007}
{Alexander}, C.~M.~{\od}., {Fogel}, M., {Yabuta}, H., \& {Cody}, G.~D. 2007,
  \gca, 71, 4380

\bibitem[{{Barnes} {et~al.}(2014){Barnes}, {Tart{\`e}se}, {Anand}, {McCubbin},
  {Franchi}, {Starkey}, \& {Russell}}]{barnes2014}
{Barnes}, J.~J., {Tart{\`e}se}, R., {Anand}, M., {McCubbin}, F.~M., {Franchi},
  I.~A., {Starkey}, N.~A., \& {Russell}, S.~S. 2014, Earth and Planetary
  Science Letters, 390, 244

\bibitem[{{Bergin} {et~al.}(2014){Bergin}, {Cleeves}, {Crockett}, \&
  {Blake}}]{bergin2014far}
{Bergin}, E.~A., {Cleeves}, L.~I., {Crockett}, N., \& {Blake}, G.~A. 2014,
  Faraday Discussions, 168, 61

\bibitem[{{Bethell} \& {Bergin}(2011{\natexlab{a}})}]{bethell2011x}
{Bethell}, T.~J. \& {Bergin}, E.~A. 2011{\natexlab{a}}, \apj, 740, 7

\bibitem[{{Bethell} \& {Bergin}(2011{\natexlab{b}})}]{bethell2011u}
---. 2011{\natexlab{b}}, \apj, 739, 78

\bibitem[{{Birnstiel} \& {Andrews}(2014)}]{birnstiel2014}
{Birnstiel}, T. \& {Andrews}, S.~M. 2014, \apj, 780, 153

\bibitem[{{Bockel{\'e}e-Morvan} {et~al.}(1998){Bockel{\'e}e-Morvan}, {Gautier},
  {Lis}, {Young}, {Keene}, {Phillips}, {Owen}, {Crovisier}, {Goldsmith},
  {Bergin}, {Despois}, \& {Wootten}}]{bockeleemorvan1998}
{Bockel{\'e}e-Morvan}, D., {Gautier}, D., {Lis}, D.~C., {Young}, K., {Keene},
  J., {Phillips}, T., {Owen}, T., {Crovisier}, J., {Goldsmith}, P.~F.,
  {Bergin}, E.~A., {Despois}, D., \& {Wootten}, A. 1998, \icarus, 133, 147

\bibitem[{{Busemann} {et~al.}(2006){Busemann}, {Young}, {Alexander}, {Hoppe},
  {Mukhopadhyay}, \& {Nittler}}]{busemann2006}
{Busemann}, H., {Young}, A.~F., {Alexander}, C.~M.~{\od}., {Hoppe}, P.,
  {Mukhopadhyay}, S., \& {Nittler}, L.~R. 2006, Science, 312, 727

\bibitem[{{Charnley} {et~al.}(1997){Charnley}, {Tielens}, \&
  {Rodgers}}]{charnley1997}
{Charnley}, S.~B., {Tielens}, A.~G.~G.~M., \& {Rodgers}, S.~D. 1997, \apjl,
  482, L203

\bibitem[{{Cleeves} {et~al.}(2013{\natexlab{a}}){Cleeves}, {Adams}, \&
  {Bergin}}]{cleeves2013a}
{Cleeves}, L.~I., {Adams}, F.~C., \& {Bergin}, E.~A. 2013{\natexlab{a}}, \apj,
  772, 5

\bibitem[{{Cleeves} {et~al.}(2013{\natexlab{b}}){Cleeves}, {Adams}, {Bergin},
  \& {Visser}}]{cleeves2013b}
{Cleeves}, L.~I., {Adams}, F.~C., {Bergin}, E.~A., \& {Visser}, R.
  2013{\natexlab{b}}, \apj, 777, 28

\bibitem[{{Cleeves} {et~al.}(2014{\natexlab{a}}){Cleeves}, {Bergin}, \&
  {Adams}}]{cleeves2014par}
{Cleeves}, L.~I., {Bergin}, E.~A., \& {Adams}, F.~C. 2014{\natexlab{a}}, \apj,
  794, 123

\bibitem[{{Cleeves} {et~al.}(2014{\natexlab{b}}){Cleeves}, {Bergin},
  {Alexander}, {Du}, {Graninger}, {{\"O}berg}, \& {Harries}}]{cleeves2014wat}
{Cleeves}, L.~I., {Bergin}, E.~A., {Alexander}, C.~M.~{\od}., {Du}, F.,
  {Graninger}, D., {{\"O}berg}, K.~I., \& {Harries}, T.~J. 2014{\natexlab{b}},
  Science, 345, 1590

\bibitem[{{Cleeves} {et~al.}(2015){Cleeves}, {Bergin}, {Qi}, {Adams}, \&
  {{\"O}berg}}]{cleeves2015tw}
{Cleeves}, L.~I., {Bergin}, E.~A., {Qi}, C., {Adams}, F.~C., \& {{\"O}berg},
  K.~I. 2015, \apj, 799, 204

\bibitem[{Coplen(1994)}]{coplen1994}
Coplen, T.~B. 1994, Pure and Applied Chemistry, 66, 273

\bibitem[{{Coutens} {et~al.}(2012){Coutens}, {Vastel}, {Caux}, {Ceccarelli},
  {Bottinelli}, {Wiesenfeld}, {Faure}, {Scribano}, \& {Kahane}}]{coutens2012}
{Coutens}, A., {Vastel}, C., {Caux}, E., {Ceccarelli}, C., {Bottinelli}, S.,
  {Wiesenfeld}, L., {Faure}, A., {Scribano}, Y., \& {Kahane}, C. 2012, \aap,
  539, A132

\bibitem[{{Dones} {et~al.}(2004){Dones}, {Weissman}, {Levison}, \&
  {Duncan}}]{dones2004}
{Dones}, L., {Weissman}, P.~R., {Levison}, H.~F., \& {Duncan}, M.~J. 2004, in
  Comets II, ed. G.~W. {Kronk}, 153--174

\bibitem[{{Eberhardt} {et~al.}(1995){Eberhardt}, {Reber}, {Krankowsky}, \&
  {Hodges}}]{eberhardt1995}
{Eberhardt}, P., {Reber}, M., {Krankowsky}, D., \& {Hodges}, R.~R. 1995, \aap,
  302, 301

\bibitem[{{Favre} {et~al.}(2015){Favre}, {Bergin}, {Cleeves}, {Hersant}, {Qi},
  \& {Aikawa}}]{favre2015}
{Favre}, C., {Bergin}, E.~A., {Cleeves}, L.~I., {Hersant}, F., {Qi}, C., \&
  {Aikawa}, Y. 2015, \apjl, 802, L23

\bibitem[{{Feuchtgruber} {et~al.}(2013){Feuchtgruber}, {Lellouch}, {Orton}, {de
  Graauw}, {Vandenbussche}, {Swinyard}, {Moreno}, {Jarchow}, {Billebaud},
  {Cavali{\'e}}, {Sidher}, \& {Hartogh}}]{feuchtgruber2013}
{Feuchtgruber}, H., {Lellouch}, E., {Orton}, G., {de Graauw}, T.,
  {Vandenbussche}, B., {Swinyard}, B., {Moreno}, R., {Jarchow}, C.,
  {Billebaud}, F., {Cavali{\'e}}, T., {Sidher}, S., \& {Hartogh}, P. 2013,
  \aap, 551, A126

\bibitem[{{Flaherty} {et~al.}(2015){Flaherty}, {Hughes}, {Rosenfeld},
  {Andrews}, {Chiang}, {Simon}, {Kerzner}, \& {Wilner}}]{flaherty2015}
{Flaherty}, K.~M., {Hughes}, A.~M., {Rosenfeld}, K.~A., {Andrews}, S.~M.,
  {Chiang}, E., {Simon}, J.~B., {Kerzner}, S., \& {Wilner}, D.~J. 2015, \apj,
  813, 99

\bibitem[{{Fogel} {et~al.}(2011){Fogel}, {Bethell}, {Bergin}, {Calvet}, \&
  {Semenov}}]{fogel2011}
{Fogel}, J.~K.~J., {Bethell}, T.~J., {Bergin}, E.~A., {Calvet}, N., \&
  {Semenov}, D. 2011, \apj, 726, 29

\bibitem[{{Furuya} \& {Aikawa}(2014)}]{furuya2014}
{Furuya}, K. \& {Aikawa}, Y. 2014, \apj, 790, 97

\bibitem[{{Furuya} {et~al.}(2013){Furuya}, {Aikawa}, {Nomura}, {Hersant}, \&
  {Wakelam}}]{furuya2013}
{Furuya}, K., {Aikawa}, Y., {Nomura}, H., {Hersant}, F., \& {Wakelam}, V. 2013,
  \apj, 779, 11

\bibitem[{{Geiss} \& {Gloeckler}(2003)}]{geiss2003}
{Geiss}, J. \& {Gloeckler}, G. 2003, \ssr, 106, 3

\bibitem[{{Gibb} {et~al.}(2004){Gibb}, {Whittet}, {Boogert}, \&
  {Tielens}}]{gibb2004}
{Gibb}, E.~L., {Whittet}, D.~C.~B., {Boogert}, A.~C.~A., \& {Tielens},
  A.~G.~G.~M. 2004, \apjs, 151, 35

\bibitem[{{Glassgold} {et~al.}(1997){Glassgold}, {Najita}, \&
  {Igea}}]{glassgold1997}
{Glassgold}, A.~E., {Najita}, J., \& {Igea}, J. 1997, \apj, 480, 344

\bibitem[{{Gourier} {et~al.}(2008){Gourier}, {Robert}, {Delpoux}, {Binet},
  {Vezin}, {Moissette}, \& {Derenne}}]{gourier2008}
{Gourier}, D., {Robert}, F., {Delpoux}, O., {Binet}, L., {Vezin}, H.,
  {Moissette}, A., \& {Derenne}, S. 2008, \gca, 72, 1914

\bibitem[{{Hartogh} {et~al.}(2011){Hartogh}, {Lis}, {Bockel{\'e}e-Morvan}, {de
  Val-Borro}, {Biver}, {K{\"u}ppers}, {Emprechtinger}, {Bergin}, {Crovisier},
  {Rengel}, {Moreno}, {Szutowicz}, \& {Blake}}]{hartogh2011}
{Hartogh}, P., {Lis}, D.~C., {Bockel{\'e}e-Morvan}, D., {de Val-Borro}, M.,
  {Biver}, N., {K{\"u}ppers}, M., {Emprechtinger}, M., {Bergin}, E.~A.,
  {Crovisier}, J., {Rengel}, M., {Moreno}, R., {Szutowicz}, S., \& {Blake},
  G.~A. 2011, \nat, 478, 218

\bibitem[{{Herrero} {et~al.}(2010){Herrero}, {G{\'a}lvez}, {Mat{\'e}}, \&
  {Escribano}}]{herrero2010}
{Herrero}, V.~J., {G{\'a}lvez}, {\accO}., {Mat{\'e}}, B., \& {Escribano}, R.
  2010, Phys. Chem. Chem. Phys., 12, 3164

\bibitem[{{Hidaka} {et~al.}(2009){Hidaka}, {Watanabe}, {Kouchi}, \&
  {Watanabe}}]{hidaka2009}
{Hidaka}, H., {Watanabe}, M., {Kouchi}, A., \& {Watanabe}, N. 2009, \apj, 702,
  291

\bibitem[{{Hugo} {et~al.}(2009){Hugo}, {Asvany}, \& {Schlemmer}}]{hugo2009}
{Hugo}, E., {Asvany}, O., \& {Schlemmer}, S. 2009, \jcp, 130, 164302

\bibitem[{{Kawakita} \& {Kobayashi}(2009)}]{kawakita2009}
{Kawakita}, H. \& {Kobayashi}, H. 2009, \apj, 693, 388

\bibitem[{{Kruijer} {et~al.}(2014){Kruijer}, {Kleine}, {Fischer-G{\"o}dde},
  {Burkhardt}, \& {Wieler}}]{kruijer2014}
{Kruijer}, T.~S., {Kleine}, T., {Fischer-G{\"o}dde}, M., {Burkhardt}, C., \&
  {Wieler}, R. 2014, Earth and Planetary Science Letters, 403, 317

\bibitem[{{Lee} \& {Bergin}(2015)}]{lee2015}
{Lee}, J.-E. \& {Bergin}, E.~A. 2015, \apj, 799, 104

\bibitem[{{Mahaffy} {et~al.}(1998){Mahaffy}, {Donahue}, {Atreya}, {Owen}, \&
  {Niemann}}]{mahaffy1998}
{Mahaffy}, P.~R., {Donahue}, T.~M., {Atreya}, S.~K., {Owen}, T.~C., \&
  {Niemann}, H.~B. 1998, \ssr, 84, 251

\bibitem[{{Meier} {et~al.}(1998){Meier}, {Owen}, {Matthews}, {Jewitt},
  {Bockelee-Morvan}, {Biver}, {Crovisier}, \& {Gautier}}]{meier1998a}
{Meier}, R., {Owen}, T.~C., {Matthews}, H.~E., {Jewitt}, D.~C.,
  {Bockelee-Morvan}, D., {Biver}, N., {Crovisier}, J., \& {Gautier}, D. 1998,
  Science, 279, 842

\bibitem[{{Messenger}(2000)}]{messenger2000}
{Messenger}, S. 2000, \nat, 404, 968

\bibitem[{{Milam} {et~al.}(2011){Milam}, {Charnley}, {Chuang}, {Kuan},
  {Coulson}, \& {Remijan}}]{milam2011}
{Milam}, S.~N., {Charnley}, S.~B., {Chuang}, Y.-L., {Kuan}, Y.-J., {Coulson},
  I.~M., \& {Remijan}, A.~R. 2011, in Lunar and Planetary Inst. Technical
  Report, Vol.~42, Lunar and Planetary Institute Science Conference Abstracts,
  1847

\bibitem[{{Millar} {et~al.}(1989){Millar}, {Bennett}, \& {Herbst}}]{millar1989}
{Millar}, T.~J., {Bennett}, A., \& {Herbst}, E. 1989, \apj, 340, 906

\bibitem[{{Mumma} \& {Charnley}(2011)}]{mumma2011}
{Mumma}, M.~J. \& {Charnley}, S.~B. 2011, \araa, 49, 471

\bibitem[{{{\"O}berg} {et~al.}(2011){{\"O}berg}, {Boogert}, {Pontoppidan}, {van
  den Broek}, {van Dishoeck}, {Bottinelli}, {Blake}, \&
  {Evans}}]{oberg2011spitz}
{{\"O}berg}, K.~I., {Boogert}, A.~C.~A., {Pontoppidan}, K.~M., {van den Broek},
  S., {van Dishoeck}, E.~F., {Bottinelli}, S., {Blake}, G.~A., \& {Evans}, II,
  N.~J. 2011, \apj, 740, 109

\bibitem[{{{\"O}berg} {et~al.}(2015){{\"O}berg}, {Furuya}, {Loomis}, {Aikawa},
  {Andrews}, {Qi}, {van Dishoeck}, \& {Wilner}}]{oberg2015d}
{{\"O}berg}, K.~I., {Furuya}, K., {Loomis}, R., {Aikawa}, Y., {Andrews}, S.~M.,
  {Qi}, C., {van Dishoeck}, E.~F., \& {Wilner}, D.~J. 2015, \apj, 810, 112

\bibitem[{{Owen} \& {Jacquet}(2015)}]{owen2015}
{Owen}, J.~E. \& {Jacquet}, E. 2015, \mnras, 446, 3285

\bibitem[{{Parise} {et~al.}(2006){Parise}, {Ceccarelli}, {Tielens}, {Castets},
  {Caux}, {Lefloch}, \& {Maret}}]{parise2006}
{Parise}, B., {Ceccarelli}, C., {Tielens}, A.~G.~G.~M., {Castets}, A., {Caux},
  E., {Lefloch}, B., \& {Maret}, S. 2006, \aap, 453, 949

\bibitem[{{Parise} {et~al.}(2003){Parise}, {Simon}, {Caux}, {Dartois},
  {Ceccarelli}, {Rayner}, \& {Tielens}}]{parise2003}
{Parise}, B., {Simon}, T., {Caux}, E., {Dartois}, E., {Ceccarelli}, C.,
  {Rayner}, J., \& {Tielens}, A.~G.~G.~M. 2003, \aap, 410, 897

\bibitem[{{Persson} {et~al.}(2012){Persson}, {J{\o}rgensen}, \& {van
  Dishoeck}}]{persson2012}
{Persson}, M.~V., {J{\o}rgensen}, J.~K., \& {van Dishoeck}, E.~F. 2012, \aap,
  541, A39

\bibitem[{{Persson} {et~al.}(2014){Persson}, {J{\o}rgensen}, {van Dishoeck}, \&
  {Harsono}}]{persson2014}
{Persson}, M.~V., {J{\o}rgensen}, J.~K., {van Dishoeck}, E.~F., \& {Harsono},
  D. 2014, \aap, 563, A74

\bibitem[{{Reboussin} {et~al.}(2015){Reboussin}, {Wakelam}, {Guilloteau},
  {Hersant}, \& {Dutrey}}]{reboussin2015}
{Reboussin}, L., {Wakelam}, V., {Guilloteau}, S., {Hersant}, F., \& {Dutrey},
  A. 2015, \aap, 579, A82

\bibitem[{{Remusat} {et~al.}(2006){Remusat}, {Palhol}, {Robert}, {Derenne}, \&
  {France-Lanord}}]{remusat2006}
{Remusat}, L., {Palhol}, F., {Robert}, F., {Derenne}, S., \& {France-Lanord},
  C. 2006, Earth and Planetary Science Letters, 243, 15

\bibitem[{{Robert} {et~al.}(2011){Robert}, {Derenne}, {Thomen}, {Anquetil}, \&
  {Hassouni}}]{robert2011}
{Robert}, F., {Derenne}, S., {Thomen}, A., {Anquetil}, C., \& {Hassouni}, K.
  2011, \gca, 75, 7522

\bibitem[{{Roberts} {et~al.}(2002){Roberts}, {Fuller}, {Millar}, {Hatchell}, \&
  {Buckle}}]{roberts2002}
{Roberts}, H., {Fuller}, G.~A., {Millar}, T.~J., {Hatchell}, J., \& {Buckle},
  J.~V. 2002, \aap, 381, 1026

\bibitem[{{Roueff} {et~al.}(2013){Roueff}, {Gerin}, {Lis}, {Wootten},
  {Marcelino}, {Cernicharo}, \& {Tercero}}]{roueff2013}
{Roueff}, E., {Gerin}, M., {Lis}, D.~C., {Wootten}, A., {Marcelino}, N.,
  {Cernicharo}, J., \& {Tercero}, B. 2013, Journal of Physical Chemistry A,
  117, 9959

\bibitem[{{Sandford} {et~al.}(2000){Sandford}, {Bernstein}, {Allamandola},
  {Gillette}, \& {Zare}}]{sandford2000}
{Sandford}, S.~A., {Bernstein}, M.~P., {Allamandola}, L.~J., {Gillette}, J.~S.,
  \& {Zare}, R.~N. 2000, \apj, 538, 691

\bibitem[{{Tielens}(1997)}]{tielens1997}
{Tielens}, A.~G.~G.~M. 1997, in American Institute of Physics Conference
  Series, Vol. 402, American Institute of Physics Conference Series, ed. E.~K.
  {Zinner} \& T.~J. {Bernatowicz}, 523--544

\bibitem[{{Turner} \& {Drake}(2009)}]{turner2009}
{Turner}, N.~J. \& {Drake}, J.~F. 2009, \apj, 703, 2152

\bibitem[{{Usui} {et~al.}(2012){Usui}, {Alexander}, {Wang}, {Simon}, \&
  {Jones}}]{usui2012}
{Usui}, T., {Alexander}, C.~M.~{\od}., {Wang}, J., {Simon}, J.~I., \& {Jones},
  J.~H. 2012, Earth and Planetary Science Letters, 357-358, 119

\bibitem[{Wakelam \& Herbst(2008)}]{wakelam2008}
Wakelam, V. \& Herbst, E. 2008, The Astrophysical Journal, 680, 371

\bibitem[{{Willacy}(2007)}]{willacy2007}
{Willacy}, K. 2007, \apj, 660, 441

\bibitem[{{Willacy} \& {Woods}(2009)}]{willacy2009}
{Willacy}, K. \& {Woods}, P.~M. 2009, \apj, 703, 479

\bibitem[{{Yang} {et~al.}(2013){Yang}, {Ciesla}, \& {Alexander}}]{yang2013}
{Yang}, L., {Ciesla}, F.~J., \& {Alexander}, C.~M.~{\od}. 2013, \icarus, 226,
  256

\end{thebibliography}
\end{document}